\documentclass[conference]{IEEEtran}

\usepackage{amsmath,amsfonts,amssymb,graphicx}
\usepackage{algorithm,algpseudocode}
\usepackage{caption}
\usepackage{subcaption}
\usepackage[mathscr]{eucal}

\usepackage{mathtools}
\usepackage[subnum]{cases}

\usepackage{float}

\usepackage{url}

\usepackage{verbatim}

\usepackage{booktabs}
\usepackage{multirow} 

\usepackage{fixltx2e}

\usepackage{flushend}

\newcommand{\argminy}{\ensuremath{\displaystyle\operatornamewithlimits{arg\,min}_y}}

\newcommand{\Code}[1]{\texttt{#1}}

\DeclareGraphicsExtensions{.pdf}
\graphicspath{img}
\title{Tolerating Silent Data Corruption in Opaque Preconditioners}

\author{
\IEEEauthorblockN{
James Elliott\IEEEauthorrefmark{1}\IEEEauthorrefmark{2},
Mark Hoemmen\IEEEauthorrefmark{2}, and
Frank Mueller\IEEEauthorrefmark{1}}
\IEEEauthorblockA{
\IEEEauthorrefmark{1}
Computer Science Department, North Carolina State University,
Raleigh, NC
}
\IEEEauthorblockA{
\IEEEauthorrefmark{2}
Sandia National Laboratories
Albuquerque, NM
}
}

\newlength{\SingleColumnAboveCaption}
\begin{document}

\maketitle

\begin{abstract}
We demonstrate algorithm-based fault tolerance for silent, transient data
corruption in ``black-box'' preconditioners.  We consider both additive Schwarz
domain decomposition with an ILU(k) subdomain solver, and algebraic multigrid,
both implemented in the Trilinos library.  We evaluate faults that corrupt
preconditioner results in both single and multiple MPI ranks.  We then analyze
how our approach behaves when then application is scaled.  Our technique is
based on a Selective Reliability approach that performs most operations in an
unreliable mode, with only a few operations performed reliably.  We also
investigate two responses to faults and discuss the performance overheads
imposed by each.  For a non-symmetric problem solved using GMRES and ILU, we
show that at scale our fault tolerance approach incurs only 22\% overhead for the
worst case. With detection techniques, we are able to reduce this
overhead to 1.8\% in the worst case.

\end{abstract}

\section{Introduction} \label{S:intro}

Recent studies indicate that large parallel computers will continue to
become less reliable as energy constraints tighten, component counts
increase, and manufacturing sizes decrease
\cite{michalak12,kogge08,geist11}.  This unreliability may manifest in
two different ways: either as ``hard'' faults, which cause the loss of
one or more parallel processes, or as ``soft'' faults, which cause
incorrect arithmetic or storage, but do not kill the running
application.  Large-scale systems today experience frequent process
loss.  Applications recover from it using global checkpoint / restart (C/R),
with current research looking at optimizing the process through, e.g.,
multi-level checkpointing \cite{moody10} or by using domain knowledge
of algorithms to create checkpoint schemes that have lower overhead
\cite{chen11}.

This paper focuses on soft faults.  Specifically, we consider those
that corrupt data or computations, without the hardware or system
detecting them and notifying the application that a fault occurred.
We call this type of soft fault Silent Data Corruption (SDC).  SDC is
much less frequent than process failures, but much more threatening,
since the application may silently return an incorrect answer.  In
physical simulations, the wrong answer could have costly and even
life-threatening consequences.  Users' trust in the results of
numerical simulations can lead to disaster if those results are wrong,
as for example in the 1991 collapse of the Sleipner A oil platform
\cite{schlaich1993ursache}.  Unlike with hard faults, applications
currently have few recovery strategies.  Hardware \emph{detection}
without correction may cost nearly as much as full hardware
correction.  Hardware vendors can harden chips against soft faults,
but doing so will increase chip complexity and likely either increase
energy usage or decrease performance.  An open field of research and
the focus of this paper is designing algorithms that can tolerate SDC.

\textbf{We present the following contributions}:
\begin{itemize} 
\item We show an algorithm-based fault tolerance approach for
  iterative linear solvers that enables the use of opaque
  preconditioners without algorithmic or code modifications.
\item We demonstrate our approach for two different iterative solvers:
  the Conjugate Gradient method (CG) for symmetric positive definite
  matrices, and the Generalized Minimal Residual Method (GMRES) for
  nonsymmetric matrices.  We do so for two preconditioners: additive
  Schwarz preconditioner with an ILU(k) subdomain solver, and
  algebraic multigrid (AMG).
\item We relate the scale at which the problem is run to the overhead
  introduced by our fault tolerance approach.
\item We present detection and rollback strategies that work with our
  fault-tolerance approach to reduce lost time due to SDC
  significantly.
\end{itemize}

Many engineering and scientific applications solve large sparse linear
systems using iterative algorithms.  Almost all of these solves use
\emph{preconditioners}.  They may take time to set up, and will add to
the time per iteration, but aim to save overall run time by reducing
the total number of iterations.  An important contribution of our work
is that we treat preconditioners as \emph{opaque}.  That is, we do not
need to modify the preconditioning algorithm or implementation.  This
matters because preconditioners are often orders of magnitude more
complex than the iterative solvers that utilize them, in terms of both
algorithms and lines of code.  For example, one can express CG in a
dozen lines of code, but the algebraic multigrid implementation MueLu
\cite{MueLu2012} we use in this work includes over half a million
lines of code from several Trilinos packages, as of the 11.8 Trilinos
release, not counting third-party sparse direct factorization
libraries for coarse-grid solves.  

Prior work in algorithm-based fault tolerance for linear algebra
problems focused on introducing checksums or other encoding schemes
into the algorithm itself.  One designs the algorithm to maintain a
checksum relationship either while it is running, or after it
completes \cite{davies11,davies13,chen08,huang84,shantharam12}.  
This requires in-depth knowledge of the method
and data structures, and requires modifying algorithms to
incorporate fault detection and correction code.  
Even if this could be done for preconditioner algorithms as
complicated as algebraic multigrid, adding checksums to an
implementation would be impractical and error-prone, given the sheer
amount of code to modify.  Such a code would need to be rewritten from
the ground up to use checksums.

Rather than attempt to rewrite every method to use an encoding scheme,
we advocate a selective reliability approach
\cite{bridges12,elliott14} that focuses fault tolerance effort on
iterative solvers rather than preconditioners.  The primary focus of
this work is the use of preconditioners in linear solvers, where the
preconditioner is assumed to be faulty. We consider the domain
decomposition technique Additive Schwarz using an Incomplete LU
factorization ILU(k), and a multi-level preconditioner Algebraic
Multigrid.  We also demonstrate that selective reliability can be used
to create other fault-tolerant solvers besides the original
``FT-GMRES'' \cite{bridges12}.

We also assess a concept presented by Elliott et al. \cite{elliott14} showing that enforcing
bounded error is sufficient for iterative linear solvers.  We consider the
``Skeptical Programming'' approach presented by Elliott et al. \cite{elliott14}, and draw
conclusions about the amalgamation of these approaches.  The faults that
motivate these fault tolerance approaches require scale to be observed.
We conjecture that any approach has to consider the implications of both strong and weak
scaling.  Specifically, we evaluate a preconditioning strategy that runs at
scale and keeps faults local, while also evaluating a preconditioning
approach that may spread corruption to other processes as part preconditioning.

\emph{This paper is organized as follows}:
\begin{enumerate}
\item In \S\ref{S:solvers}, we introduce the preconditioned linear
  solvers we evaluate.
\item In \S\ref{S:preconditioners}, we describe the types of
  preconditioners we chose and discuss how they propagate errors.
\item In \S\ref{S:fault_model}, we describe our fault injection
  methodology, and explain how we characterize faults.
\item In \S\ref{S:results}, we present findings that show the average
  number of extra preconditioner calls, given various types of faults.
\item In \S\ref{S:conclusion}, we summarize results and discuss future
  work.
\end{enumerate}

\section{Preconditioned Linear Solvers} \label{S:solvers}

We consider two solvers that are capable of solving different classes
of problems. The Generalized Minimal Residual Method (GMRES) from Saad
and Schultz \cite{saad86} can solve nonsymmetric problems.  The Method
of Conjugate Gradients (CG) \cite{hestenes52} can only solve symmetric
positive definite (SPD) linear systems, but is faster for doing so.
CG is used in the NAS Parallel Benchmarks \cite{bailey91} and in
Mantevo miniapps like HPCCG \cite{heroux2009improving}.

Linear solvers often utilize preconditioners as a means to accelerate
convergence.  More specifically, preconditioning is a transformation
that attempts to improve some aspect of the linear system, typically
the condition number.  We consider preconditioning in two different
ways.  Left preconditioning solves the preconditioned problem $M^{-1}
(Ax-b)$, which requires a preconditioner application at the start of
the solve.  Right preconditioning solves $AM^{-1}Mx = b$, and requires
a preconditioner application to also compute a solution update.

Algorithm~\ref{alg:GMRES} presents right-preconditioned GMRES.  GMRES
applies the preconditioner in lines~\ref{alg:gmres:PCApply} and
\ref{alg:gmres:SolutionUpdatePCApply}.  Note that the solution update
(Line~\ref{alg:gmres:SolutionUpdatePCApply}) need not be calculated
every iteration, and is often not computed until the solver exits.
\begin{algorithm}[htp]
\caption{(Right-preconditioned) GMRES}
\label{alg:GMRES}
\begin{algorithmic}[1]
\Input{Linear system $Ax=b$ and initial guess $x_0$}
\Output{Approximate solution $x_m$ for some $m \geq 0$}
\State{$r_0 := b - A x_0$}\Comment{Unpreconditioned initial residual vector}
\State{$\beta := \| r_0 \|_2$}
\State{$q_1 := r_0 / \beta$}
\For{$j = 1, 2, \dots$ until convergence}
  \State{Solve $M z_j = q_j$ for $z_j$}\Comment{Apply preconditioner}
    \label{alg:gmres:PCApply}
  \State{$v_{j+1} := A z_j$}\Comment{Apply the matrix $A$}
  \For{$i = 1, 2, \dots, k$}\Comment{Orthogonalize}
    \State{$H(i,j) := q_i^* v_{j+1}$}
     \label{alg:gmres:upperHess}
    \State{$v_{j+1} := v_{j+1} - q_i H(i,j)$}
  \EndFor
  \State{$H(j+1,j) := \| v_{j+1} \|_2$}
  \State{$q_{j+1} := v_{j+1} / H(j+1,j)$}\Comment{New basis vector}
  \State{$y_j := \argminy \| H(1:j+1,1:j) y - \beta e_1 \|_2$}
  \State{$x_j := x_0 + M^{-1} [q_1, q_2, \dots, q_j]
  y_j$}\Comment{Solution update}
    \label{alg:gmres:SolutionUpdatePCApply}
\EndFor
\end{algorithmic}
\end{algorithm}
Algorithm~\ref{alg:CG} presents left-preconditioned CG.
Preconditioner applications occur on Lines \ref{alg:cg:InitialPCApply}
and \ref{alg:cg:PCApply}.  Note that left-preconditioned CG does not
require a preconditioner application to compute its solution update.
\begin{algorithm}[htp]
\caption{(Left-preconditioned) CG}
\label{alg:CG}
\begin{algorithmic}[1]
\Input{Linear system $Ax=b$ and initial guess $x_0$}
\Output{Approximate solution $x_m$ for some $m \geq 0$}
\State{$r_0 := b - A x_0$}\Comment{Unpreconditioned initial residual
vector}
\State{Solve $M z_0 = r_0$ for $z_0$}\Comment{Apply preconditioner}
    \label{alg:cg:InitialPCApply}
\State{$p_0 := z_0$}
\For{$j = 1, 2, \dots$ until convergence}
  \State{$\alpha_j := (r_j,z_j)/(Ap_j,p_j)$}
  
  \State{$x_{j+1} := x_j + \alpha_j p_j$}
  
  \State{$r_{j+1} := r_j - \alpha_j Ap_j$}
  
  \State{Solve $M z_{j+1} = r_{j+1}$ for $z_{j+1}$}\Comment{Apply
  preconditioner}
    \label{alg:cg:PCApply}
  
  \State{$\beta_j := (r_{j+1},z_{j+1})/(r_j,z_j)$}
  
  \State{$p_{j+1} := z_{j+1} + \beta_j p_j$}
\EndFor
\end{algorithmic}
\end{algorithm}

Our results center around observing the number of extra preconditioner
applications relative to solving the problem without SDC.  That is, we
observe the impact of $\tilde{z} = M^{-1} w$, where $\tilde{z}$
indicates the corrupted output of a preconditioner call. In
\S~\ref{S:preconditioners} we give more details on on how we decompose
problems across multiple processes, and in \S~\ref{S:fault_model} we
explain what $\tilde{z}$ looks like given faults on some (or all)
parallel processes.

\subsection{Selective Reliability}

Our fault-tolerance strategy rests on relating numerical methods that
naturally correct errors to system-level fault tolerance.  In
particular, we assume a \emph{selective reliability} or ``sandboxing''
programming model that lets algorithm developers isolate faults to
certain parts of the algorithm in a coarse-grained way.  Bridges et
al.~\cite{bridges12} used this idea to develop the fault-tolerant
linear solver FT-GMRES.  FT-GMRES uses a reliable ``outer'' solver,
preconditioned by an unreliable ``inner'' solver.  Any faults that
occur in the inner solver manifest as a ``different preconditioner''
to the outer solver.  The outer solver is chosen to be Flexible GMRES
\cite{saad03}, which can tolerate a preconditioner that changes
between iterations. The inner solver was chosen to be GMRES, though
any linear solver would work.  Nested solvers are a common idea in
numerical algorithms; the authors applied this idea to fault tolerance
by choosing the right outer solver and observing that inner solves
could run unreliably.  Selective reliability is a programming model
that requires codesign between algorithms, system software, and
hardware.  Neither Bridges et al.~\cite{bridges12} nor we attempt to implement this
programming model; possible implementation strategies include
redundancy or software error-correcting codes.

In this paper, we express FT-GMRES as \Code{FGmres->Gmres}. This
notation captures the outer solver \Code{FGmres} and the inner solver
\Code{Gmres}.  Should the inner solver uses a preconditioner, e.g.,
\Code{ILU}, we then write \Code{FGmres->Gmres->ILU}.  The authors of
FT-GMRES assumed that since GMRES is considered ``robust'' as a
solver, it would makes sense to use \Code{FGmres->Gmres} over
combinations like \Code{FGmres->Cg}.  We evaluate this in
\S~\ref{S:results}.

\subsection{Implementation}

We implemented our solvers using the Tpetra \cite{Tpetra} sparse linear algebra
package in the Trilinos framework \cite{heroux03} and validated them
against both MATLAB and the solvers in Trilinos' Belos package \cite{Amesos2Belos}.
Implementing our solvers using Trilinos lets us benefit from the
scalability and performance of its sparse matrices and dense vectors.
In addition, basing our research in Trilinos also gives us access to a
wealth of numerical algorithms, on
which we elaborate in \S~\ref{S:preconditioners}.

\section{Preconditioners} \label{S:preconditioners}

This paper shows how solvers behave in the presence of faulty
preconditioners.  Given that our solvers are parallel, we must
consider parallel preconditioners.  We examine two popular examples:
single level additive Schwarz domain decomposition with no overlap,
and Algebraic Multigrid (AMG).  We have chosen preconditioners that
utilize these decomposition techniques because each exhibits different
types of communication patterns with respect to preconditioning.

\subsection{Additive Schwarz}

Single-level additive Schwarz domain decomposition (see \cite{smith96})
divides the solution vector into $k$ subdomains, either with or without overlap.
Typical MPI implementations assign one subdomain to each MPI process.
Each subdomain applies a solver (of any kind) locally.  
Additive Schwarz then ``glues'' the subdomains' results back together to
form the preconditioner's output.  We use Trilinos' implementation in the
Ifpack2 package to create our preconditioner, and choose
Incomplete LU with zero fill, \Code{ILU(0)} (see \cite{saad03}), 
as our subdomain solver.

We use no overlap with additive Schwarz, because this means that any corruption in a
subdomain will not impact output from other subdomains. That is, with zero
overlap, there is no communication between subdomains after applying the local
preconditioner.  This differs from multigrid preconditioners.

\subsection{Algebraic Multigrid}
Algebraic Multigrid (AMG) is a robust multilevel preconditioner.  While
geometric Multigrid requires knowledge of a grid, AMG operates directly on the
matrix.  In a setup phase, restriction operators are defined that ``coarsen''
the matrix, creating consecutively smaller matrices. Likewise, prolongation
operators are determined that interpolate the coarse level information back  to
finer levels of the multigrid hierarchy.  Coarsening from the finest level to
coarsest and back is referred to as a V-cycle.  Prior to prolongation, AMG
applies a smoother to the current level.  Smoothers are often cheap solvers,
e.g., a single sweep of Jacobi or Gauss-Seidel.  We choose a single Gauss-Seidel
sweep at all but the coarsest level, and we use SuperLU \cite{superlu99} as the smoother for our
coarsest matrix.

\subsection{Hierarchy vs. Overlap}
\begin{figure}[htp]
  \centering
  \begin{subfigure}[t]{0.5\columnwidth}
    \setlength{\abovecaptionskip}{\SingleColumnAboveCaption}
    \centering
    \includegraphics[width=0.8\columnwidth]{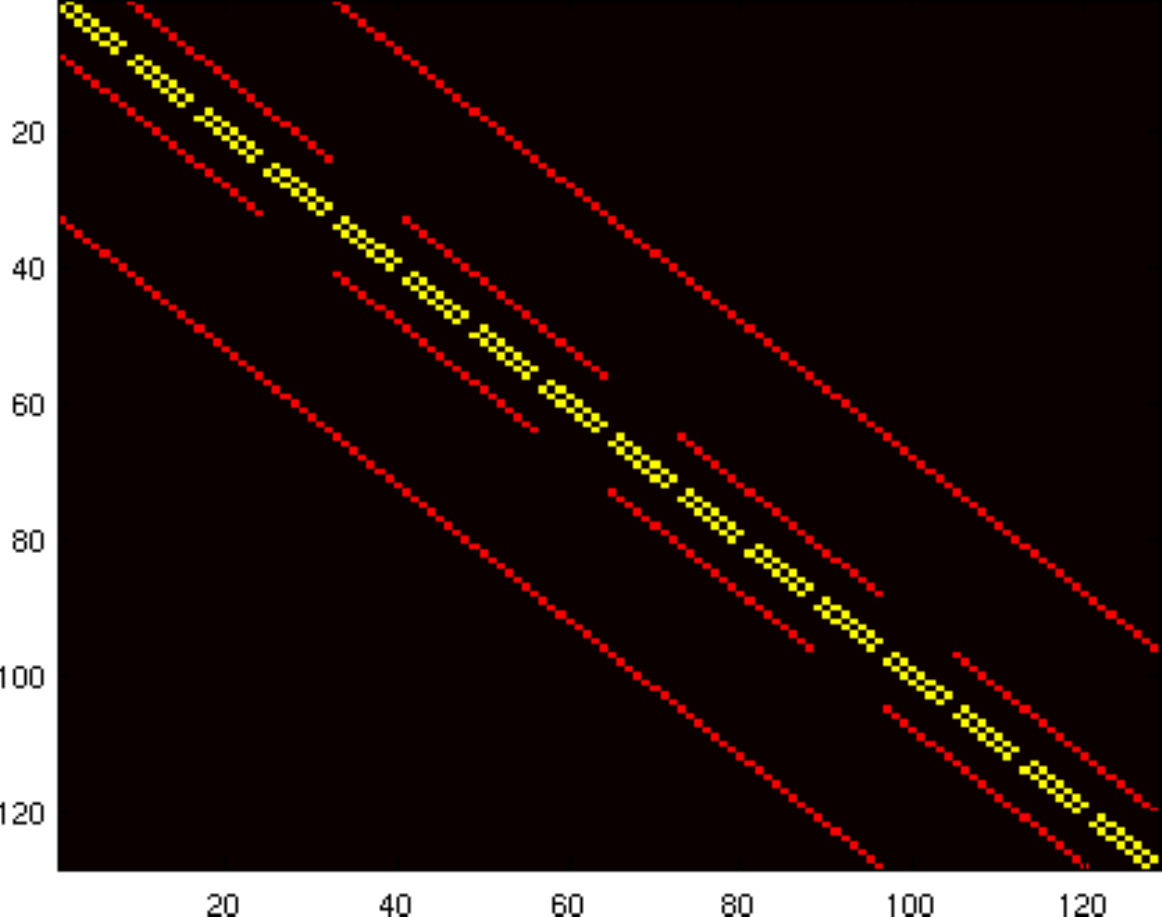}
    \caption{Finest}
    \label{F:comm_pattern:muelu:finest}
  \end{subfigure}%
  \begin{subfigure}[t]{0.5\columnwidth}
    \setlength{\abovecaptionskip}{\SingleColumnAboveCaption}
    \centering
    \includegraphics[width=0.8\columnwidth]{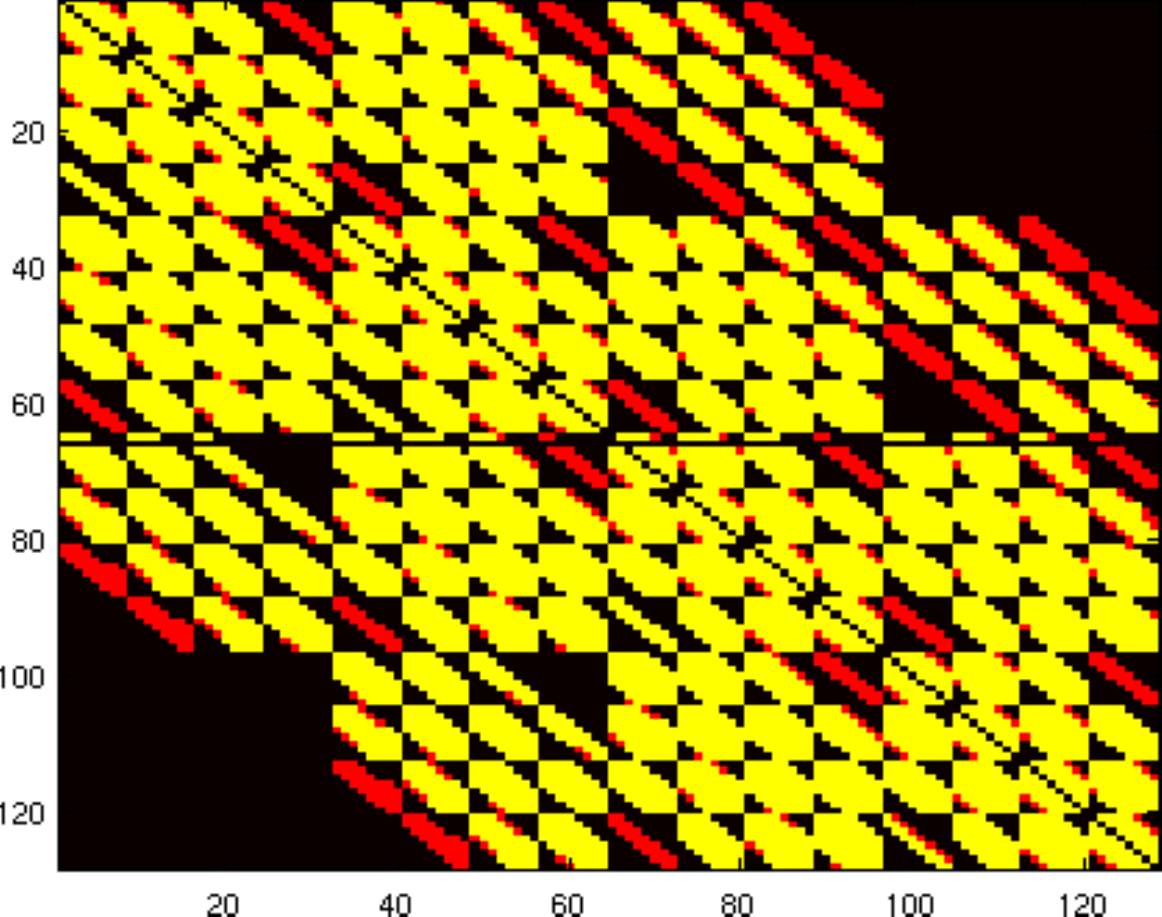}
    \caption{Coarsest}
    \label{F:comm_pattern:muelu:coarsest}
  \end{subfigure}%
\setlength{\abovecaptionskip}{0.5\baselineskip}
\setlength{\belowcaptionskip}{-0.5\baselineskip}
  \caption{Communication patterns for MueLu at different hierarchy levels
  (no repartitioning).}
  \label{F:comm_pattern:muelu}
\end{figure}

Due to AMG's hierarchical structure, a fault in a multigrid method may
propagate from the process where the fault occurs to other processes.  Should SDC
occur at the coarsest level, it is possible that half (or all) of the nodes
absorb some amount of corruption into their final solutions.  Alternatively, if
SDC occurs in one process' data at the finest level, the error will remain local if no further
V-cycles are performed.  (In our setup, with MueLu as a preconditioner, we
enforce only one V-cycle.)  To illustrate this communication pattern, we 
show the communication pattern at the finest and coarsest levels for a 7 level
AMG configuration in Fig.~\ref{F:comm_pattern:muelu}.  
This differs from additive Schwarz: without overlap, subdomains do not
communicate, and even with overlap, they only communicate with their
nearest neighbors.

\section{Fault Model and Injection Methodology} \label{S:fault_model} 

This work is motivated by the premise that soft errors will become
more likely in future systems, and that SDC has been observed in
current systems.  Much uncertainty remains about how how often soft
errors will occur, and how they will manifest in applications.  Most
prior work modeled soft errors as one or more flipped bits e.g., \cite{sloan12}.
Researchers injected bit flips and observed their effects on running
applications \cite{shantharam11,davies13,sao13}.  These works
showed that current algorithms \emph{can} misbehave badly if their data are corrupted.  In the
presence of SDC, iterative methods can either return the wrong answer,
or fail to converge and iterate forever.  What previous results have
\emph{not} achieved, though, is to give algorithm designers tools to
prove that one algorithm is \emph{better} than another algorithm.

For this reason, we intentionally choose a fault model that is
abstract.  Regardless of how a soft error occurs, we do know that if
the soft error creates SDC, then we can represent the SDC as a numerical
error in our algorithm.  We make neither a claim to address all problems that
soft errors may introduce, nor do we specifically focus on the event that
incorrect data are used in a calculation.  This enables our work to
address a larger set of errors than the bit flip model does, e.g.,
corruption of a network packet.  Since our results show what
happens if an entire MPI process returns tainted data, they may also
guide development of alternatives to C/R recovery that replace data
from a lost process with a best guess.

We intentionally avoid injecting faults at a high rate, because it is
not clear that soft error rates will be high.  Vendors have strong
incentives to sell reliable machines, at least at small scales of
parallelism, but may accept that the largest-scale parallel machines
might expose some faults to applications.  Alternately, some systems
might provide less reliable ``stochastic circuits'' as low-power or
high-performance accelerators, to which users may allocate portions of
their computation.  Our model covers all of these cases.

\subsection{Modeling ``Bad'' Faults in Preconditioners}

Since we do not know the answers to many questions regarding soft
errors in future systems, we choose to create errors that we know are ``very
bad'' and observe how our algorithms handle these egregious errors.  We
define ``bad'' errors in the next paragraph, but the idea is for such
errors to have meaningful effects on the solver.  We denote the vector
output of a preconditioner as $z$ (see, e.g., Line~\ref{alg:cg:PCApply}
of Algorithm~\ref{alg:CG}).  In numerical analysis, we describe errors
using norms.  GMRES minimizes the residual error with respect to the
L-2 norm, and CG minimizes the $A$-norm of the solution error, which
is also an L-2 norm of a different vector.  For this reason, we choose
to characterize errors with respect to the L-2 norm.  This leads us to
two classifications of errors: Those that change the L-2 norm
of the output, and those that preserve its L-2 norm.

We define a \emph{bad} error to be a \emph{faulty subdomain} permuting
its portion of the global vector.  \emph{Permutations} preserve the
L-2 norm.  We also consider the case that this bad error changes the
L-2 norm. To \emph{change} the L-2 norm, a faulty subdomain may also
\emph{scale} its portion of the vector by some constant (faulty domains
\emph{always} permute).
Scaling a portion of the global (distributed) vector might not change the L-2
norm by that amount globally, since only a portion of the vector is
scaled. Should we fault \emph{all} ranks, then we see in that case
that the global vector's length is changed by some factor. 

\subsection{Granularity of Faults}

In our fault model, we let subdomains (one subdomain per MPI
process) return completely corrupt solutions.  That is, we consider faults
at the MPI process level, rather than single
values.  This seems pessimistic, but it lets us model faults 
\emph{inside} a preconditioner that may affect more than one entry 
of its output vector, and possibly even more than one process.
For example, an incorrect pivot in a sparse factorization for AMG's
coarse-grid solve may cause incorrect values on \emph{all} processes.
What our fault model promises is to characterize SDC that arises from
the preconditioner in the inner solver of our fault-tolerant
inner-outer iteration.  We show in our results that these types of
faults are sufficiently bad to cause the entire inner solve to become
divergent.

\subsection{SDC and Solvers}

In FT-GMRES, GMRES was chosen as the inner solver because it is
commonly accepted as ``more robust'' than CG.  Given that CG can only
solve SPD linear systems, if a fault occurs in CG, the error can cause
problems by appearing to be nonsymmetric \cite{meek13}, and CG can behave
very poorly.  It is for these reason we that CG is not considered as
an outer solver.  SDC may also change the sign of key values, e.g., a
negative projection length.  We consider sign-changing SDC by
negatively scaling the output vector, which is length preserving if
the scaling factor is $-1$.

\section{Results} \label{S:results}
\newcommand*{\columnWidthFactor}{0.7}
\graphicspath{img/perc,img}

\subsection{Methodology}
We described in \S~\ref{S:fault_model} how we corrupt the preconditioner's
output.  To evaluate the impact of our preconditioned solvers in the presence of
SDC, we perform the following steps:
\begin{enumerate}
\item Solve the problem injecting no SDC, and compute the number of
  times, $K$, the preconditioner was applied.
  \label{S:results:method:failurefree}
\item For all $j$ in $[1,K]$, reattempt the solve, introducing SDC at
  the $j$-th preconditioner application.  This results in $K$ total
  solves.
  \label{S:results:method:failures}
\item For all $K$ solves with SDC, compute the relative percent  of
  \emph{additional} preconditioner applies over the SDC-free 
  solve\footnote{If $Applies_{observed} - Applies_{FailureFree} <
  0$, i.e., SDC accelerated convergence, we record zero overhead.}, e.g.,
  $(Applies_{observed} - Applies_{FailureFree}) / Applies_{FailureFree}
  \times 100$
  \label{S:results:method:overhead}
\item Repeat Steps \ref{S:results:method:failures} and
  \ref{S:results:method:overhead}, letting various numbers of MPI
  processes participate in the SDC injection.
  \label{S:results:method:faulty_ranks}
\item Repeat Steps
  \ref{S:results:method:failures}-\ref{S:results:method:faulty_ranks},
  varying the scaling factor applied to the SDC.
\item For each combination of scaling factor and number of faulty
  processes, plot the \emph{average} number of additional
  preconditioner applies as a percentage.  0\% means no additional
  applies; 100\% means twice as many.
\end{enumerate}

\subsection{Strong Scaling}
We present two studies. First, we fix the problem size and strongly scale by increasing the number of MPI processes.
We start with 32 processes, and then use 1032 processes on the NERSC Hopper cluster. 
We choose 5 fixed numbers of faulty ranks: 1, 2, 8, 16, and 32.
By strong scaling the problem,
 the percentage of work per process decreases.
Hence, the percent of the global output vector that is tainted by SDC decreases.  

\subsection{Weak Scaling}
Second, we weak scale a problem such that the work per process remains
fixed at $10^5$ unknowns.  In this experiment, we consider faults as a
\emph{percentage}, rather than a fixed number, of the process count.
This decision is based on the fact that strong scaling demonstrates how
increasing the process count minimizes the amount of corruption a single process can introduce. 
With weak scaling, the global problem size must increase sufficiently to
minimize the impact of a single process faulting.
Also, multilevel preconditioners communicate information at different
grid levels.  A fault at the coarsest grid could propagate, tainting data on
all nodes by the time the finest grid is reached.  Thus, it makes sense for multigrid
to explore a percentage of ranks faulting.  100\% of ranks faulting is akin
to a drastic fault at the coarsest level that corrupts everything.  50\% of ranks
faulting would be a drastic fault at the 2\textsuperscript{nd} coarsest level of
the hierarchy.

\subsection{Test Problems}
We evaluate two classes of problems and use solvers and preconditioners
appropriate for each.
We consider a SPD linear system (Poisson problem), which is solvable by both CG
and GMRES.  For this type of problem AMG (MueLu) represents a good
preconditioner.  This yields the fault tolerant solvers \Code{FGmres->Cg->MueLu}
and \Code{FGmres->Gmres->MueLu}.  We also consider a non-symmetric problem,
CoupCons3D \cite{davis11}, which is very ill-conditioned.  We may only solve
CoupCons3D using GMRES, and ILU(0) is a suitable preconditioner, yielding
\Code{FGmres->Gmres->ILU}.  We may only weak scale our Poisson problem, as it is
generated, allowing us to dynamically increase the matrix size.

\subsection{Preconditioner Effectiveness}
We intentionally chose the preconditioners evaluated.  For the problems
presented, MueLu represents a very \emph{good} preconditioner, while ILU
represents a ``better than nothing'' preconditioner. In terms of computational
cost, MueLu is much more expensive to apply than ILU, and also requires more
communication.  The increased work performed by MueLu results in it solving the
Poisson problem in approximately 6 total applies. Contrasted to ILU
and the CoupCons3D problem, \Code{FGmres->Gmres->ILU} requires 100 ILU
applications.  So, MueLu is expensive to apply, meaning
we do not want to apply it significantly more times than necessary.
Conversely, ILU is cheap to apply and additional applications are not as
costly.

\section{Strong Scaling Results} \label{S:strong_scaling}
\subsection{Incomplete LU Preconditioning}
\begin{figure}[htp]
  \centering
  \begin{subfigure}[t]{\columnwidth}
    \setlength{\abovecaptionskip}{\SingleColumnAboveCaption}
    \centering
    \includegraphics[width=\columnWidthFactor\columnwidth]{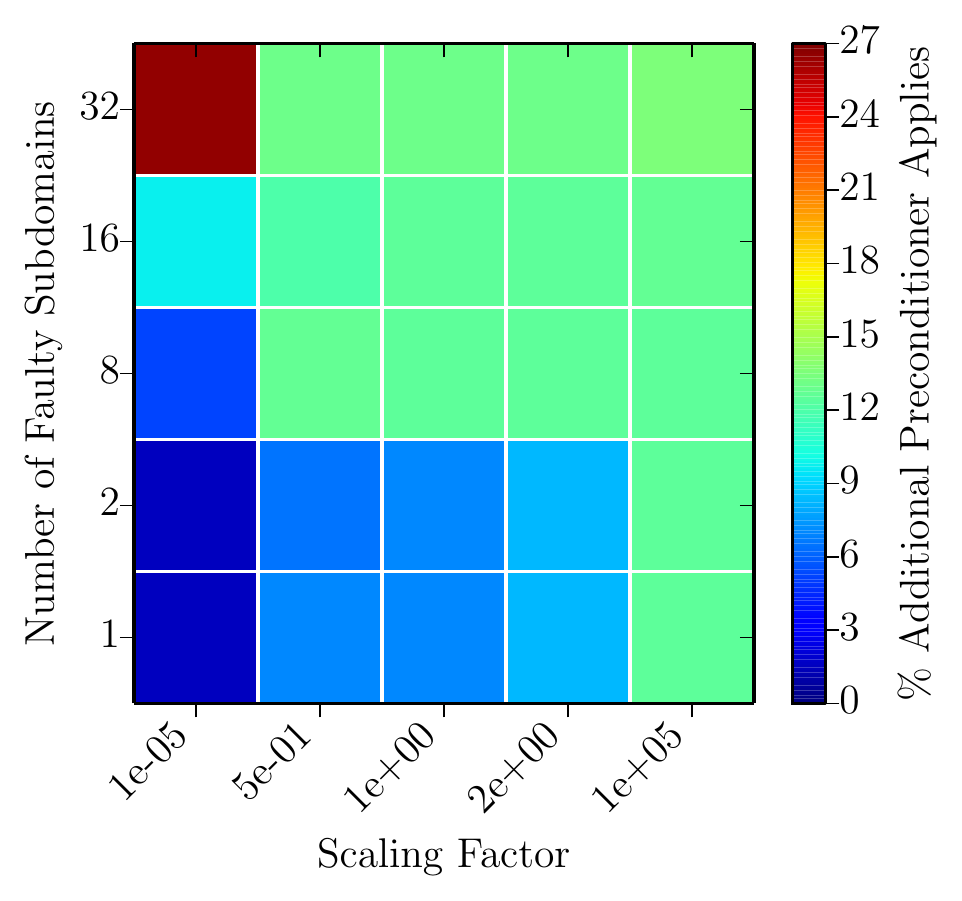}
    \caption{32 subdomains}
    \label{F:strong_scaling:gmres:CoupCons3D:32}
  \end{subfigure}%
  \\%
  \begin{subfigure}[t]{\columnwidth}
    \setlength{\abovecaptionskip}{\SingleColumnAboveCaption}
    \centering
    \includegraphics[width=\columnWidthFactor\columnwidth]{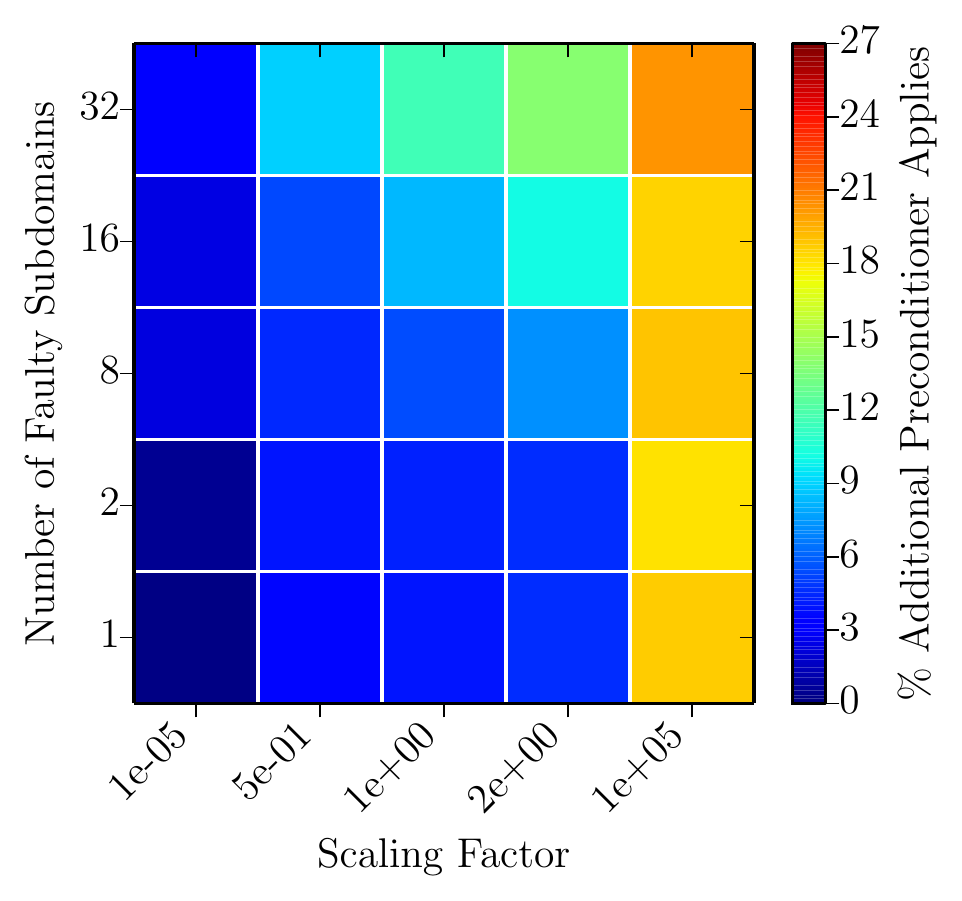}
    \caption{1032 subdomains}
    \label{F:strong_scaling:gmres:CoupCons3D:1032}
  \end{subfigure}%
\setlength{\abovecaptionskip}{0.5\baselineskip}
\setlength{\belowcaptionskip}{-0.5\baselineskip}
  \caption{Strong Scaling (32 vs. 1032 subdomains): Average percent
  increase of preconditioner applies when solving a non-symmetric problem using \Code{FGmres->Gmres->ILU} preconditioning.}
  \label{F:strong_scaling:gmres:CoupCons3D}
\end{figure}

Fig.~\ref{F:strong_scaling:gmres:CoupCons3D} shows the result of strong
scaling the CoupCons3D problem from 32 to 1032 MPI processes. For the CoupCons3D
problem, we require \textbf{100 preconditioner applications in a failure free
environment}.
Consider Fig.~\ref{F:strong_scaling:gmres:CoupCons3D:32}, the y-axis represents
the number of Additive Schwarz subdomains that participate in the fault.  The
x-axis indicates whether the SDC decreases, maintains (center column), or
increases the L-2 norm of the preconditioners output.  The color indicates the
percent increase in preconditioner applications. The bottom row of colored
squares represent rare, bad SDC. Moving vertically, we increase the number of
ranks that experience a fault.

In general, we see that a single faulty subdomain is tolerated with very low
overhead, that is, we perform little or no additional work relative to if no
fault had occurred.  We also see that SDC that increases the 2-norm is
universally bad, while SDC that maintains or shrinks the 2-norm is typically
better than increasing the 2-norm. The exception to this is the upper left-hand
block, which corresponds to 32 fault subdomains (all ranks faulted), and all
ranks decreased the 2-norm by a factor of $10^{-5}$.

In
Fig.~\ref{F:strong_scaling:gmres:CoupCons3D:1032}, we strong scale the problem
from 32 processes to 1032.  We see a trend: faults that corrupt, while
preserving or shrinking the 2-norm have considerably lower overhead than faults
that increase the 2-norm.

We can conclude from this experiment that our Selective Reliability scheme
offers dynamic fault tolerance. That is, when faults are rare, we perform less
fault tolerance work, and when faults are more common we automatically
perform more work without the need to explicitly detect and correct any errors. 
This dynamic approach is enabled by coupling systems fault tolerance and
numerical analysis.

\subsection{Algebraic Multigrid Preconditioning}
Next, we consider solving the SPD Poisson problem.  This problem is suitable for
CG or GMRES, and MueLu is a very effective preconditioner.  The effectiveness of
MueLu as a preconditioner means that we apply it a very small number of times:
We require \textbf{6 preconditioner applications in a failure free
environment}.

\begin{figure}[htp]
  \centering
  \begin{subfigure}[t]{\columnwidth}
    \setlength{\abovecaptionskip}{\SingleColumnAboveCaption}
    \centering
    \includegraphics[width=\columnWidthFactor\columnwidth]{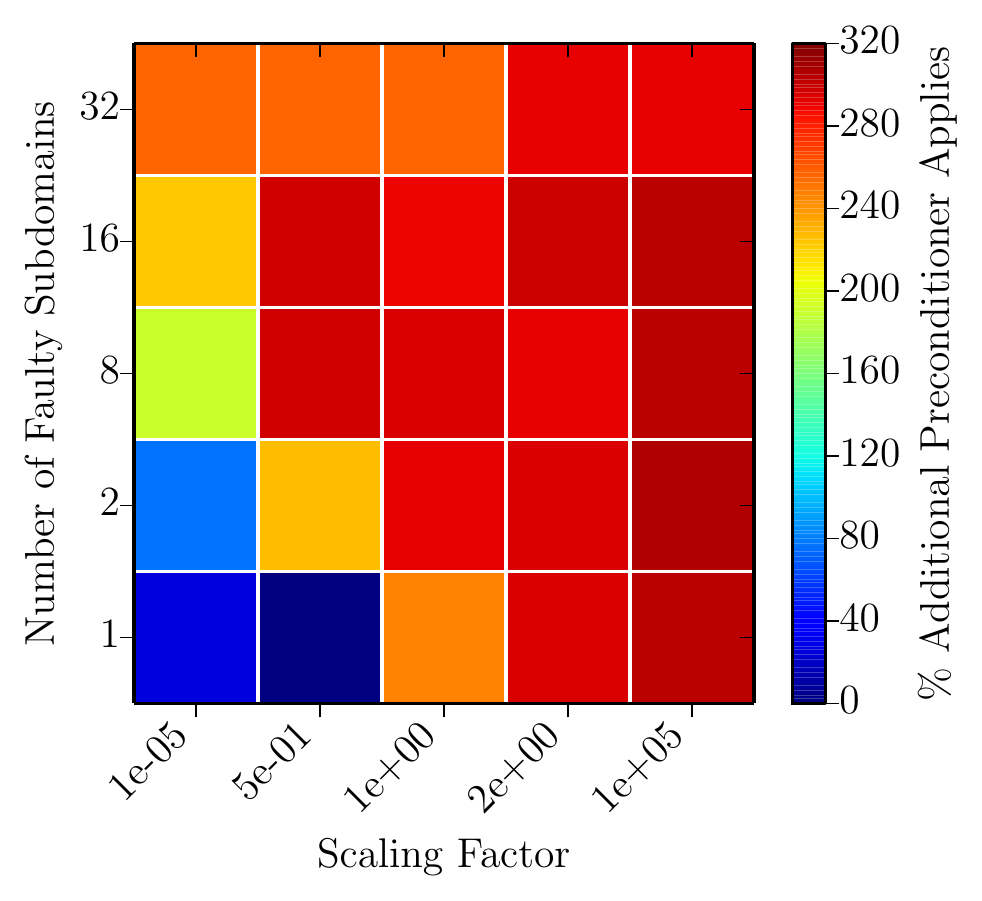}
    \caption{32 subdomains}
    \label{F:strong_scaling:cg:TpetraScalingProblem:32}
  \end{subfigure}%
  \\%
  \begin{subfigure}[t]{\columnwidth}
    \setlength{\abovecaptionskip}{\SingleColumnAboveCaption}
    \centering
    \includegraphics[width=\columnWidthFactor\columnwidth]{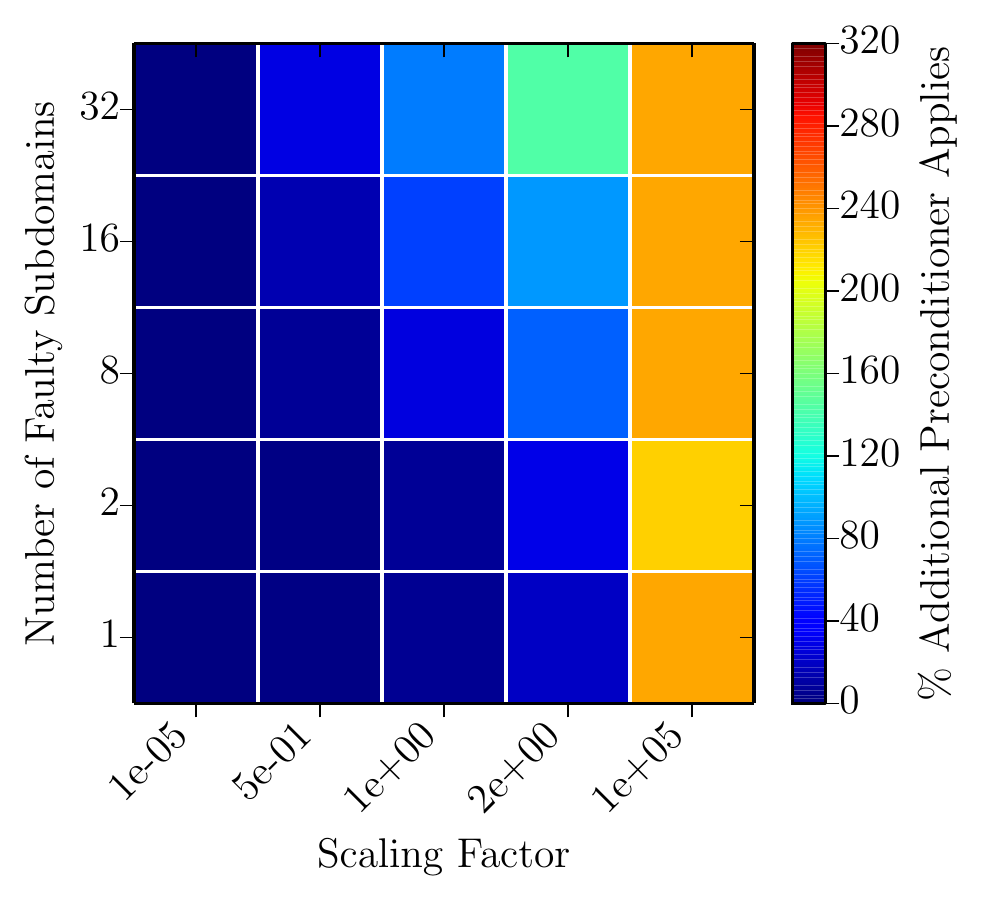}
    \caption{1032 subdomains}
    \label{F:strong_scaling:cg:TpetraScalingProblem:1032}
  \end{subfigure}%
\setlength{\abovecaptionskip}{0.5\baselineskip}
\setlength{\belowcaptionskip}{-0.5\baselineskip}
  \caption{Strong Scaling (32 vs. 1032 subdomains): Average percent
  increase of preconditioner applies when solving a symmetric positive definite problem using \Code{FGmres->Cg->MueLu} preconditioning.}
  \label{F:strong_scaling:cg:TpetraScalingProblem}
\end{figure}

Fig.~\ref{F:strong_scaling:cg:TpetraScalingProblem} shows the average percentage
of additional MueLu applies given \Code{Fgmres->Cg->MueLu} as the solver.  It is
immediately clear that there is a  trend (see
Fig.~\ref{F:strong_scaling:gmres:CoupCons3D}), where length preserving or
shrinking faults incur lower overhead than length increasing SDC.
Starting in
the lower left-hand corner of
Fig.~\ref{F:strong_scaling:cg:TpetraScalingProblem:32}, we see that SDC
induced very little additional work (typically 0 or 1 additional MueLu apply).
Moving towards the right, we observe that SDC slightly shrinks the L-2 norm of the
preconditioner output. Again, we see a low overhead. As we continue to move
towards the lower right-hand corner, we observe that SDCs introduce large overhead.
These faults correspond to length (L-2 norm) increasing SDC.  If we start in the
lower left-hand corner and move vertically, we increase the number of subdomains
that participate in the SDC. Given a single faulty subdomain (MPI
rank), we observe that
the overhead introduced by SDC is lower, and as the number of faulty subdomains
increases, the overhead from SDC increases.  This result is expected, since
increasing the number of faulty subdomains increases the amount of corruption
introduced into the calculations. And because we have strongly scaled
the experiments, this also means
we have increased the percentage of corruption. For the 32 subdomain run
(Fig.~\ref{F:strong_scaling:cg:TpetraScalingProblem:32}), the top row
shows that
\emph{every rank} faulted.

We now observe the \emph{same trend} as above, in the 1032 subdomain job.
Starting in the lower left-hand corner and moving towards the right, we see very
low overhead until we encounter an SDC that \emph{increases} the L-2 norm (bottom
right-most two squares). Here, SDC caused slightly more overhead for a small
increase in the L-2 norm, and very high overhead for a large increase (the
bottom right-most square). Moving vertically from the lower left-hand corner, we
observe an SDC that shrinks the L-2 norm also provides the lowest
overhead. When moving towards the right, the overhead increases.

We further observe a 3.2x increase
in preconditioner calls in the worst case when 32 processors are used. But when
strongly scaled to 1032 processors, we see that length preserving or decreasing
faults incur relatively low overhead, e.g., a 0-40\% increase in preconditioner
calls (0-3 additional applies).

\begin{figure}[htp]
  \centering
  \begin{subfigure}[t]{\columnwidth}
    \setlength{\abovecaptionskip}{\SingleColumnAboveCaption}
    \centering
    \includegraphics[width=\columnWidthFactor\columnwidth]{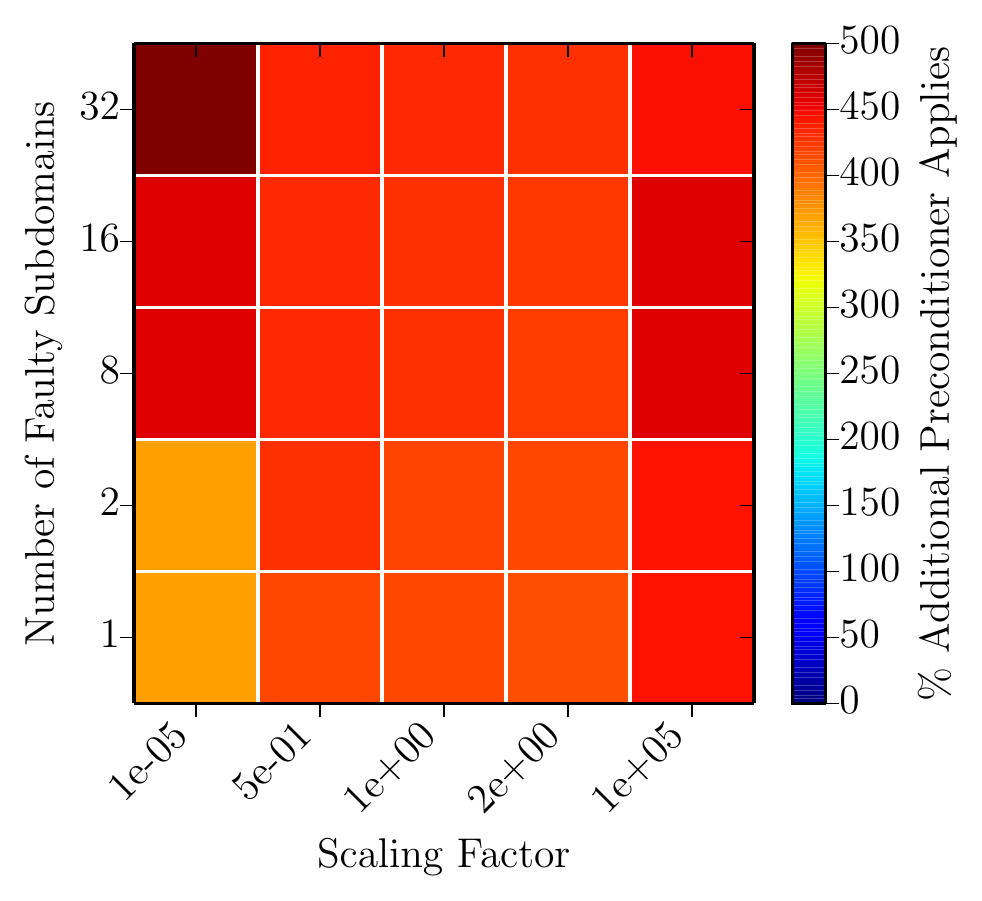}
    \caption{32 subdomains}
    \label{F:strong_scaling:gmres:TpetraScalingProblem:32}
  \end{subfigure}%
  \\%
  \begin{subfigure}[t]{\columnwidth}
    \setlength{\abovecaptionskip}{\SingleColumnAboveCaption}
    \centering
    \includegraphics[width=\columnWidthFactor\columnwidth]{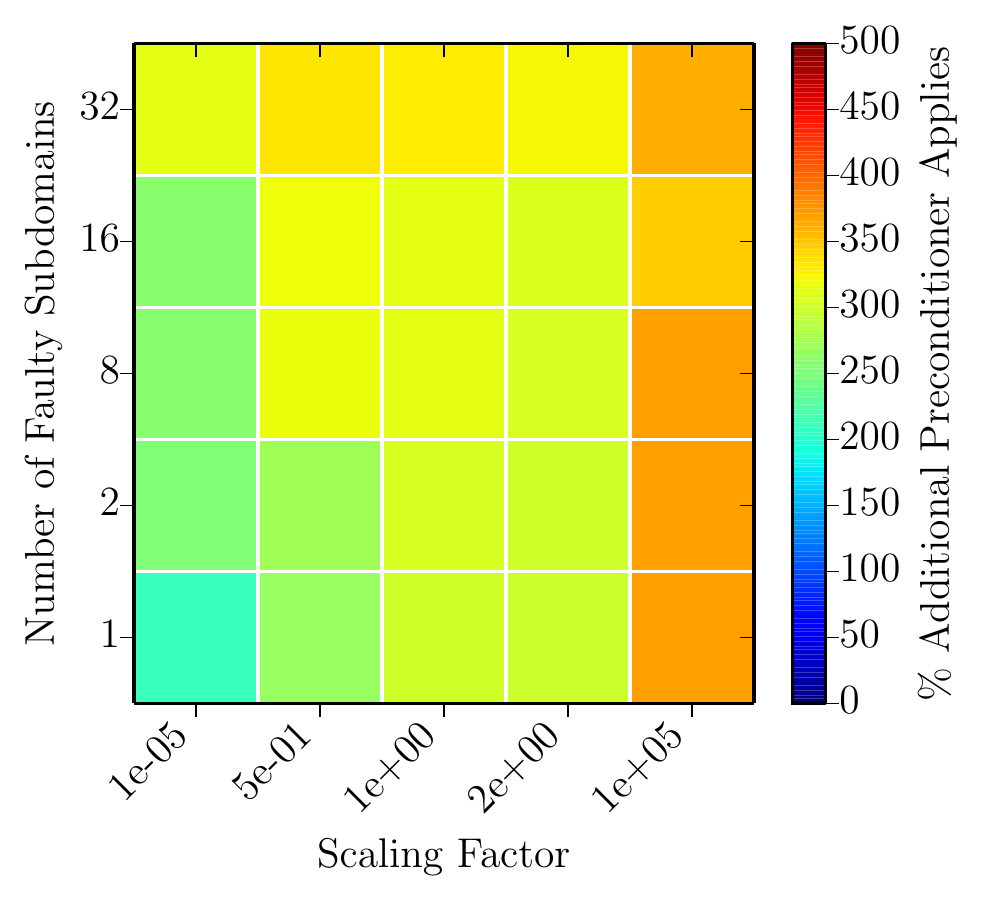}
    \caption{1032 subdomains}
    \label{F:strong_scaling:gmres:TpetraScalingProblem:1032}
  \end{subfigure}%
\setlength{\abovecaptionskip}{0.5\baselineskip}
\setlength{\belowcaptionskip}{-0.5\baselineskip}
  \caption{Strong Scaling (32 vs. 1032 subdomains): Average percent
  increase of preconditioner applies when solving a symmetric positive definite problem using \Code{FGmres->Gmres->MueLu}
  preconditioning.}
  \label{F:strong_scaling:gmres:TpetraScalingProblem}
\end{figure}

Fig.~\ref{F:strong_scaling:gmres:TpetraScalingProblem} depicts results
from \Code{FGmres->Gmres->MueLu} to solve the SPD problem.  We see the
\emph{same} trend as described in the \Code{FGmres->Cg->MueLu} case.  The lower
left-hand corner represents the lowest overhead, and moving vertically or
horizontally from this quadrant increases the overhead introduced by SDC.  This
trend is shown in both the 32 subdomain and 1032 subdomain runs.

It is immediately clear
that \Code{FGmres->Cg->MueLu} is superior in terms of overhead at both 32 and
1032 processes. The worst case for GMRES is a 500\% increase versus CG's 320\%. 
We believe the cause for this is due to GMRES preserving state (having
``memory''), while CG is
mostly stateless.  A fault in GMRES is embedded into the subspace that is built,
where as CG is effectively a 3-term recurrence.  This gives CG as an inner
solver a chance to recover without wasting an entire inner solve.  GMRES appears
to have difficulty recovering from corruption to its basis.  We leave further
analysis to future work but recommend \Code{FGmres->Cg}  for SPD problems with a good
preconditioner.  We next
consider some detection strategies, i.e., if you are using an expensive
preconditioner, it may pay off to check the explicit residual.

\subsection{Optional Detection Strategies}

Faults in inner solves are silent.  Our solvers must either converge
through them or detect them and possibly roll back a failed inner
solve.  
Detection cannot catch all SDC, and it also does not even need to, particularly
for small errors. We discussed a trend in our results showing that SDCs that
shrink or maintain the L-2 norm of the preconditioner's output are optimal
with respect to SDC that increases the L-2 norm.
Nonetheless, detection may help avoid wasted work in a failed inner solve.
In the previous section's results, we did not attempt to detect SDC or roll back
iterations in inner solves.  With an expensive but effective preconditioner,
failed inner solves may have high overhead.

In this section, we evaluate two different error detectors by
observing whether they would have triggered given the faults
presented.  We saw in the last section (compare Figures
\ref{F:strong_scaling:cg:TpetraScalingProblem} and
\ref{F:strong_scaling:gmres:TpetraScalingProblem}) that GMRES as an
inner solver has less resilience to faults than CG for SPD problems.
However, GMRES has a built-in invariant that CG lacks, namely that the
2-norm of the explicitly computed residual $r_k = b - Ax_k$ will never
increase.  Both CG and GMRES can use projection length bounds on
intermediate basis vectors to detect faults, but only GMRES can use
the monotonicity of $\|r_k\|_2$.  We now consider if detection
strategies in GMRES make sense relative to CG's relatively lower fault
tolerance overhead.

\subsection{Detection Strategies: Non-Symmetric Problem}
\begin{figure*}[htp]
  \centering
  \begin{subfigure}[t]{\columnwidth}
    \centering
    \includegraphics[width=\columnWidthFactor\columnwidth]{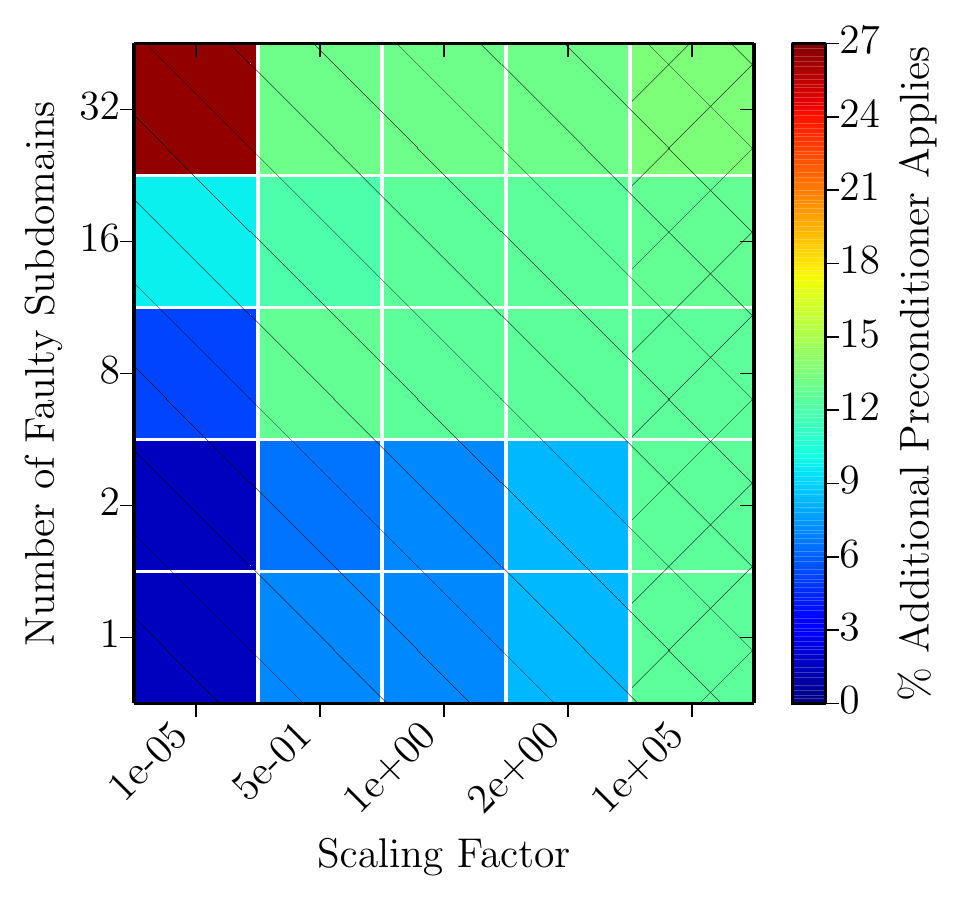}
    \caption{Faults that could be detected by the norm bound
    presented in 
    \cite{elliott14} are hatched right $/$, and faults that could
    be detected by checking the explicit residual are hatched left \textbackslash.}
    \label{F:strong_scaling:gmres:CoupCons3D:detectable}
  \end{subfigure}%
  \quad%
  \begin{subfigure}[t]{\columnwidth}
    \centering
    \includegraphics[width=\columnWidthFactor\columnwidth]{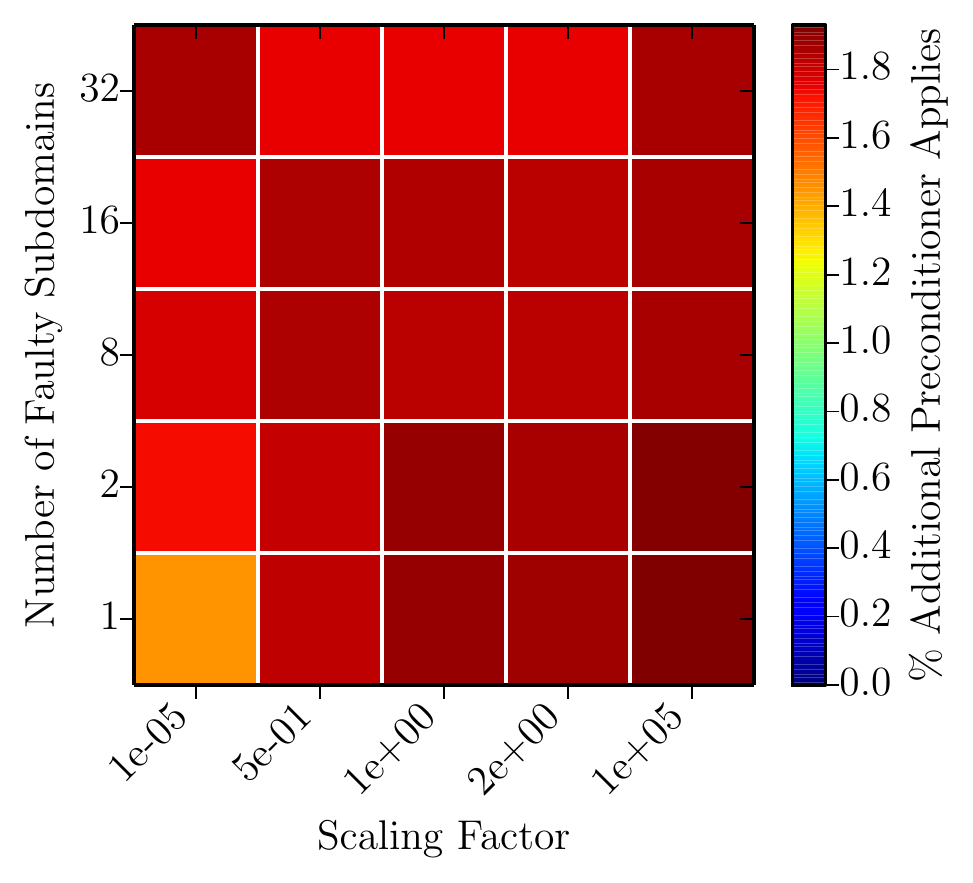}
    \caption{Responding to norm bound or explicit residual check to exit
    inner solve early (note the colorbar range is different).}
    \label{F:strong_scaling:gmres:CoupCons3D:aborting}
  \end{subfigure}%
\setlength{\abovecaptionskip}{0.5\baselineskip}
\setlength{\belowcaptionskip}{-0.5\baselineskip}
  \caption{Strong Scaling: Average percent increase of preconditioner
  applies when solving a non-symmetric problem using \Code{FGmres->Gmres->ILU} preconditioning, and checking
  projection lengths and explicit residuals every iteration. These plots uses
  32 subdomains (MPI ranks).}
  \label{F:strong_scaling:gmres:CoupCons3D:detection}
\end{figure*}

We use the data from Fig.~\ref{F:strong_scaling:gmres:CoupCons3D:32} to create
Fig.~\ref{F:strong_scaling:gmres:CoupCons3D:detectable}. We have superimposed
'\textbackslash' to indicate that the explicit residual detected the error, and
'$/$' to indicate whether a projection length norm bound proposed by Elliott et
al.~\cite{elliott14} would be triggered.  We see that the explicit residual
caught all SDCs introduced, while the projection length bound only caught errors
that drastically increased the 2-norm of the preconditioner's output.  The norm
bound proves extremely effective, given its cheapness to evaluate relative to
the cost of computing the explicit residual.

Fig.~\ref{F:strong_scaling:gmres:CoupCons3D:aborting} shows the effect
of responding to these checks.  We see at most $2 = 100 \times .018$
extra preconditioner applications.  Recall that CoupCons3D with GMRES
required 100 preconditioner applications with no faults.  Given the
cost of computing the explicit residual, relative to the cost of
applying ILU, it is not clear that checking the explicit residual
every iteration is worth the overhead.  We found experimentally that
checking the explicit residual every 5 iterations sufficed (data
omitted due to space).

\subsection{Detection Strategies: Symmetric Problem}
\begin{figure*}[htpb]
  \centering
  \begin{subfigure}[t]{\columnwidth}
    \centering
    \includegraphics[width=\columnWidthFactor\columnwidth]{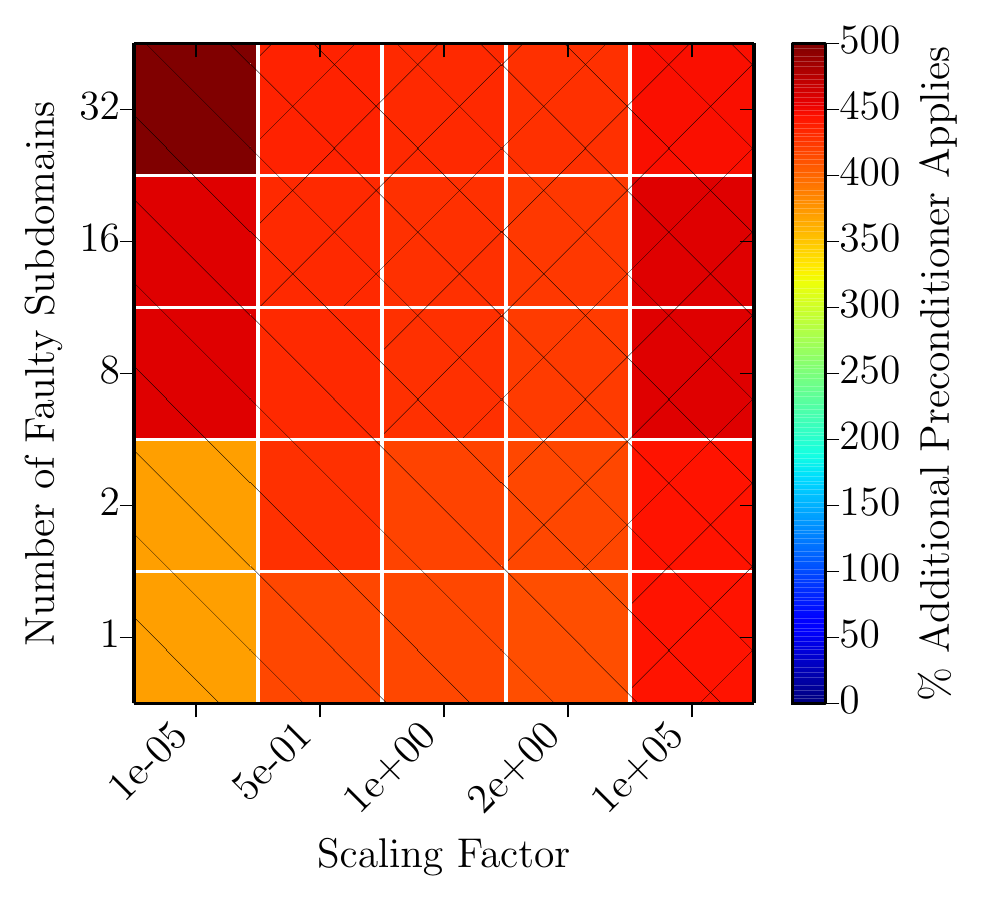}
    \caption{Faults that could be detected by the norm bound
    presented in 
    \cite{elliott14} are hatched right $/$, and faults that could
    be detected by checking the explicit residual are hatched left \textbackslash.}
    \label{F:strong_scaling:gmres:TpetraScalingProblem:detectable}
  \end{subfigure}%
  \quad%
  \begin{subfigure}[t]{\columnwidth}
    \centering
    \includegraphics[width=\columnWidthFactor\columnwidth]{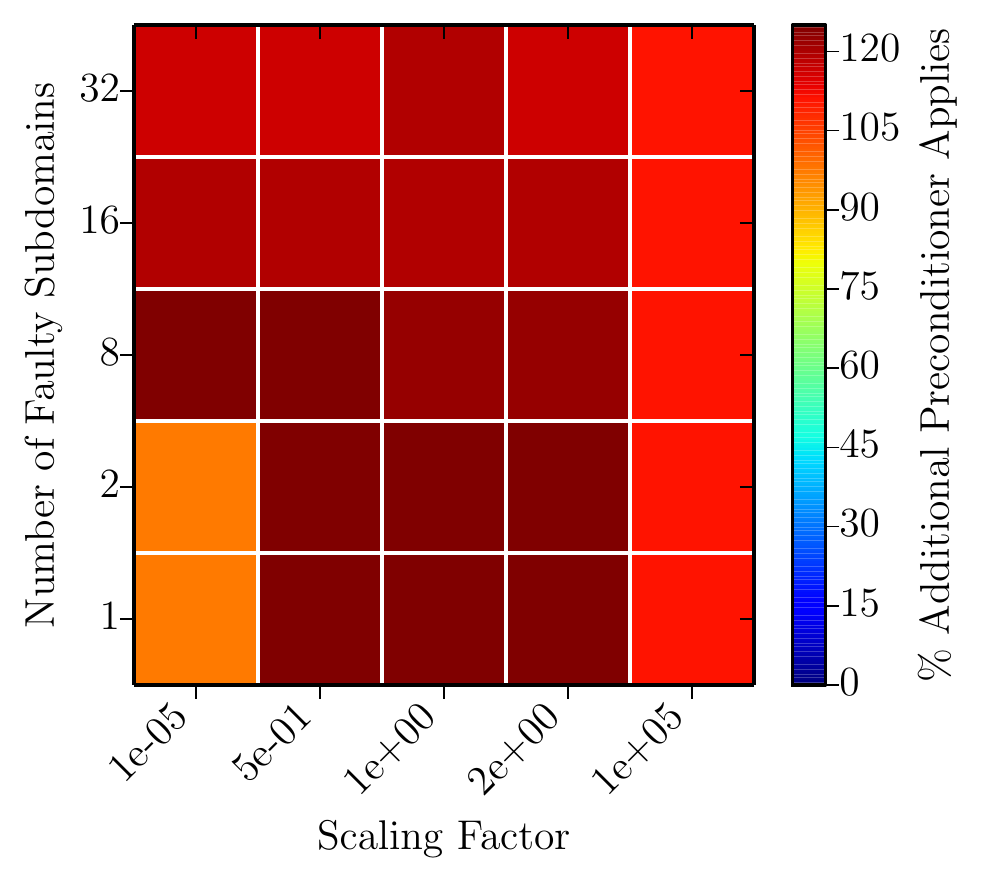}
    \caption{Responding to norm bound or explicit residual check to exit
    inner solve early (note the colorbar range is different).}
    \label{F:strong_scaling:gmres:TpetraScalingProblem:aborting}
  \end{subfigure}%
\setlength{\abovecaptionskip}{0.5\baselineskip}
\setlength{\belowcaptionskip}{-0.5\baselineskip}
  \caption{Strong Scaling: Average percent increase of preconditioner
  applies when solving an SPD problem using \Code{FGmres->Gmres->MueLu}
  preconditioning, and checking projection lengths and explicit residuals every
  iteration. These plots uses 32 subdomains (MPI ranks).}
  \label{F:strong_scaling:gmres:TpetraScalingProblem:detection}
\end{figure*}

Fig.~\ref{F:strong_scaling:gmres:TpetraScalingProblem:detection} takes the data
from Fig.~\ref{F:strong_scaling:gmres:TpetraScalingProblem:32} and superimposes
the detectors. Recall that GMRES had very poor behavior compared to CG in this
case.
We see that the norm bound by Elliott et al.~\cite{elliott14} has good coverage
for the Poisson problem. This is due to the spectral norm of A being relatively
small compared to the CoupCons3D spectral norm (which has order $10^{+6}$).  We
see that by using detectors, we can make GMRES's fault tolerance overhead much
smaller than CG: 120\% (GMRES) vs 320\% (CG). Seven extra preconditioner calls
are required for GMRES and 20 for CG.  Given the cost of applying MueLu, it pays
to use GMRES and then to compute the explicit residual.

We now take into account the cost of a GMRES iteration vs a CG iteration. 
GMRES's work grows linearly with the number of iterations it performs. We
consider work to be the number of \emph{dot products} per iteration, since
sparse matrix-vector multiplication (SpMVs) are the same for CG and GMRES.
Given that GMRES needs a total of $6+7$ inner iterations to converge (one
preconditioner apply per inner iteration) and CG required $6+20$ inner
iterations, the dot products for GMRES are $(13\times14)/2 = 91 $, while CG
requires $26\times2= 52$.

We now account for the preconditioner applications.  Right preconditioned GMRES
requires a preconditioner application to compute the explicit residual, meaning
that if the explicit residual was checked every iteration, then 13
inner iterations would require $13\times 2 = 26$ preconditioner applies. Left
preconditioned CG requires no preconditioner applications to compute the
explicit residual, and, hence, just 26 preconditioner applications suffice.

In summary, with detectors, GMRES required $91$ dot products and $26$
preconditioner applications, while CG required $52$ dot products and $26$
preconditioner applications.  These findings still favor CG as the inner solver.
 The norm bound from Elliott et al.~\cite{elliott14} flagged 50\% of the errors,
meaning no explicit residual check would be required.  We found experimentally
that checking the explicit residual every 5 iterations was sufficient for
lowering the overhead in GMRES.  Taking this into account, we can slightly
reduce the number of preconditioner applies required, but also incur more dot
products (as we can iterate up to 5 iterations before detecting that a fault
occurred).

This is a surprising result.  GMRES + detectors beats CG, but the overhead of
computing the detectors allows CG to win. If the preconditioner is sufficiently
expensive to apply, then it makes sense to use GMRES + detectors, since it can
detect and rollback the inner solve should an error occur.
CG does not support the explicit residual check, as it does not promise a
non-increasing residual.  We leave to future work incorporating the norm bound
check into CG. With a projection length check, we  expect CG to be a clear
winner over GMRES.

\subsubsection{FGMRES as the Inner Solver}
\begin{figure}[htpb]
  \centering
  \includegraphics[width=\columnWidthFactor\columnwidth]{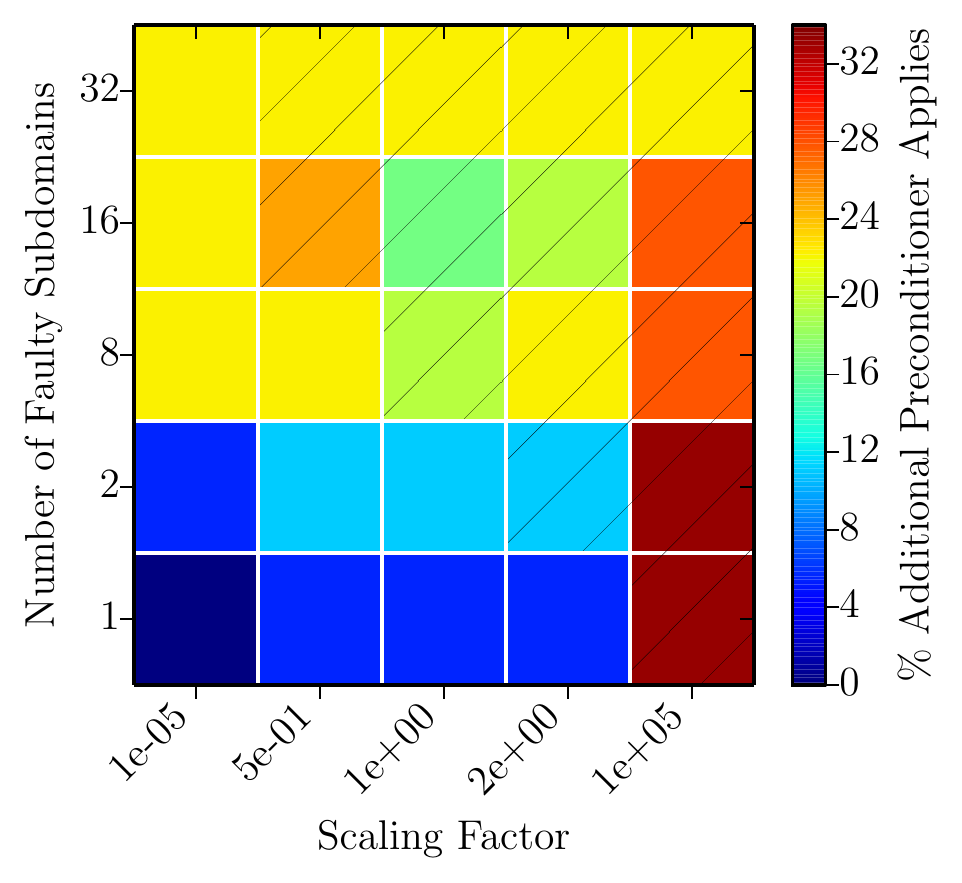}
\setlength{\abovecaptionskip}{0.5\baselineskip}
\setlength{\belowcaptionskip}{-0.5\baselineskip}
  \caption{Faults that could be detected by the norm bound presented in
  \cite{elliott14} are hatched right $/$, and faults that could
  be detected by checking the explicit residual are hatched left
  \textbackslash, when solving an SPD problem using
  \Code{FGmres->FGmres->MueLu}. This plot uses a total of 32
  subdomains (MPI ranks).}
  \label{F:strong_scaling:fgmres:TpetraScalingProblem:detectable}
\end{figure}

The premise behind Selective Reliability and nesting solvers is that users who
are \emph{not} experts in numerical algorithms can take existing solver /
preconditioner combinations and easily nest them inside a reliable Flexible
GMRES outer solver.  This enables a ``black box'' approach to iterative solver
reliability.  If we also choose the inner solver to yield the lowest fault
tolerance overhead, then we should consider all solvers, including Flexible
GMRES, as the inner solver, e.g., \Code{FGmres->FGmres->MueLu}.
Fig.~\ref{F:strong_scaling:fgmres:TpetraScalingProblem:detectable} presents this
case.  We see that if we trade memory (Flexible GMRES requires storing two sets
of basis vectors), then we obtain significantly lower fault tolerance overhead. 
Also, FGMRES does \emph{not} require the application of the preconditioner to
compute the explicit residual, meaning that explicit residual checks are
inexpensive even if the preconditioner is very expensive.

\subsubsection{Robustness of Projection Length}
\begin{figure}[htpb]
\centering
  \centering
  \includegraphics[width=\columnWidthFactor\columnwidth]{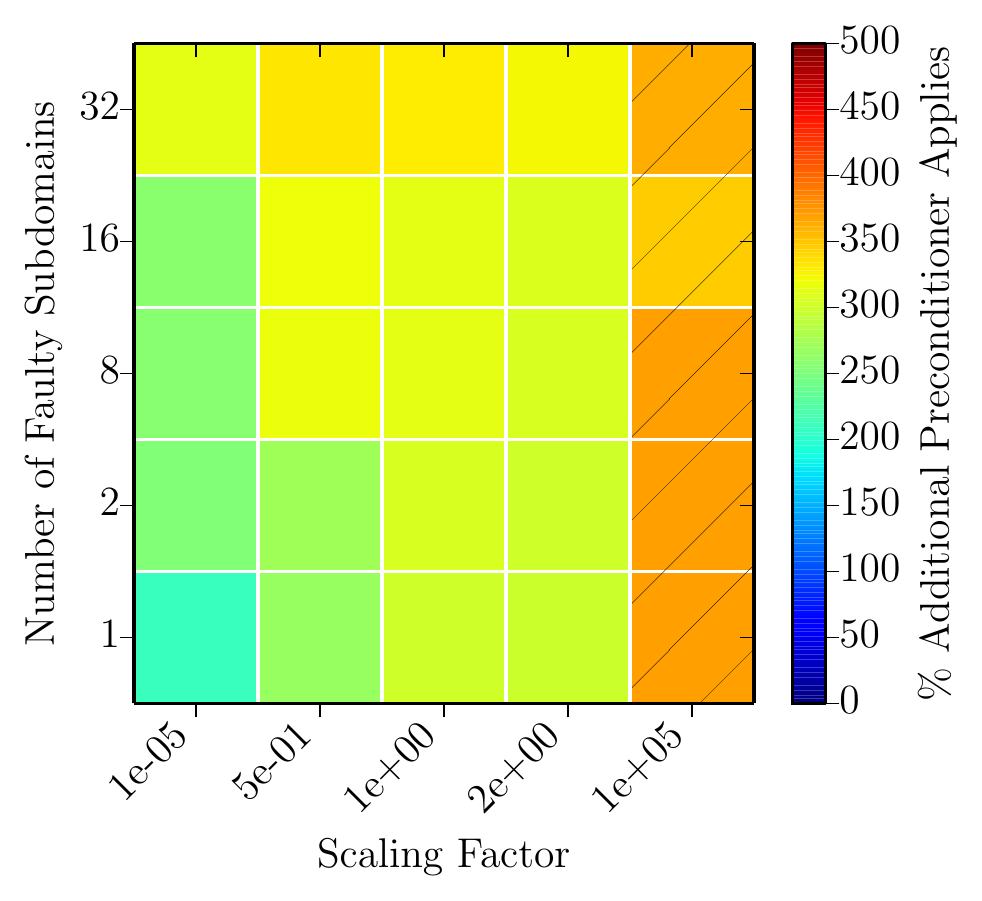}
\setlength{\abovecaptionskip}{0.5\baselineskip}
\setlength{\belowcaptionskip}{-0.5\baselineskip}
  \caption{Strong Scaling: Faults that could be detected by the norm bound
  presented in \cite{elliott14} are hatched right $/$, and faults that could
  be detected by checking the explicit residual are hatched left
  \textbackslash, when solving an SPD problem using
  \Code{FGmres->Gmres->MueLu}. This plot uses a total of 1032
  subdomains (MPI ranks) and should be compared to
  Fig.~\ref{F:strong_scaling:gmres:CoupCons3D:detectable} .}
  \label{F:strong_scaling:gmres:TpetraScalingProblem:detectable:1032}
\end{figure}

An unexpected result is shown in
Fig.~\ref{F:strong_scaling:gmres:TpetraScalingProblem:detectable}.
The projection length check flags SDC where the L-2 norm of the preconditioner
output preserves, increase, or decreases.  In this case, the norm bound from
Elliott~\cite{elliott14} is approximately $3.6\times10^{2}$, which is relatively
tight.  The projection length bound is not checked immediately after calling the
preconditioner, but is instead checked as the upper Hessenberg is formed, e.g.,
in Line~\ref{alg:gmres:upperHess} of Algorithm~\ref{alg:GMRES}.  In
Fig.~\ref{F:strong_scaling:gmres:TpetraScalingProblem:detectable:1032} we show
what is detectable when then problem is strongly scaled, e.g.,
Fig.~\ref{F:strong_scaling:gmres:TpetraScalingProblem} with detectors overlaid.
We see that the explicit residual failed to detect any errors, yet the norm
bound still succeeded in detecting the most costly errors.

\subsection{Impact of Sign Changes}

\begin{figure}[htp]
  \centering
  \includegraphics[width=\columnwidth]{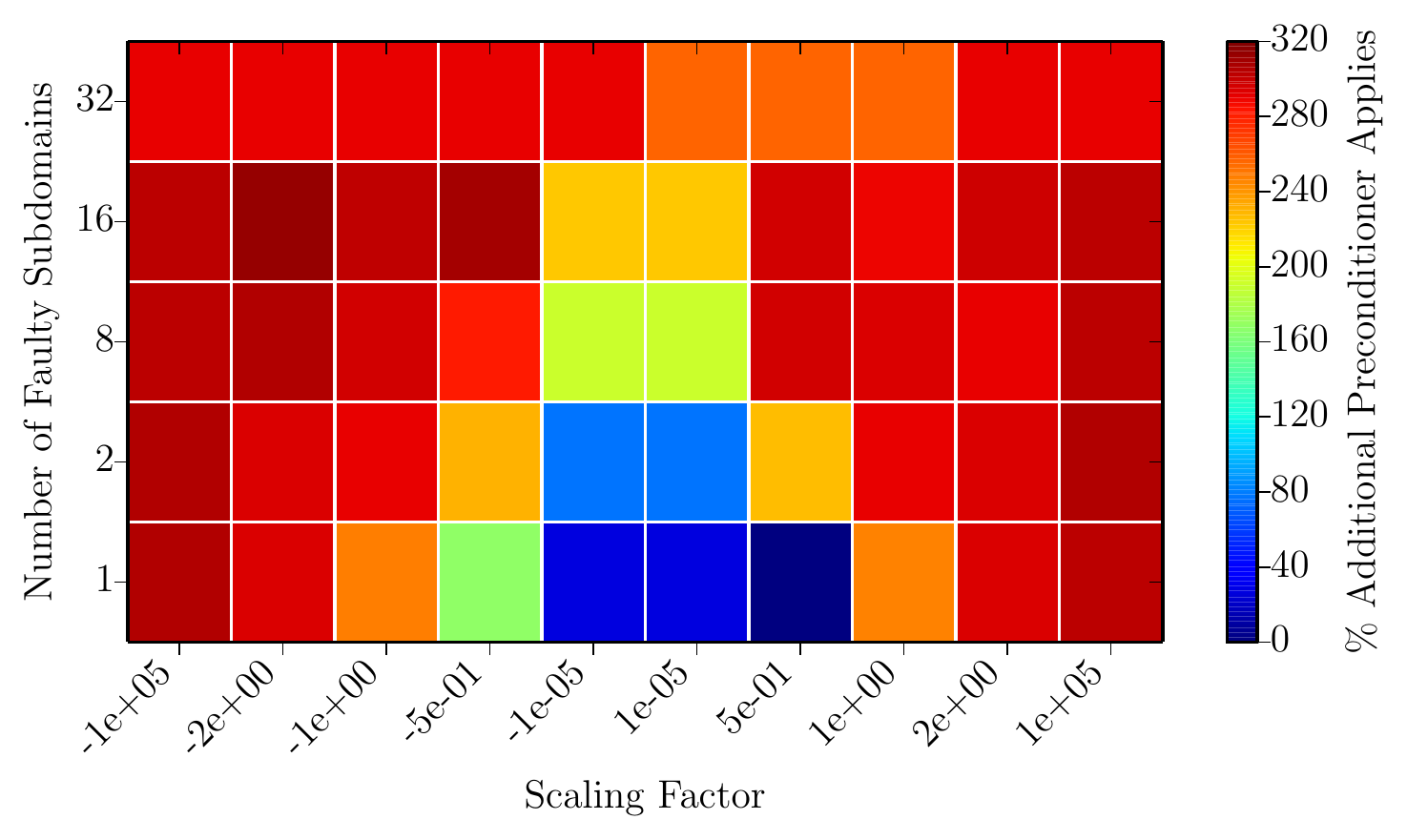}
  \caption{Strong Scaling: Average number of additional preconditioner
    applies given faults that change the sign of the preconditioner
    output when solving with \Code{FGmres->Cg->MueLu}.  This plot uses
    32 subdomains (MPI processes).}
  \label{F:strong_scaling:neg_pos:cg}
\end{figure}

We now consider the impact of a sign-changing fault with CG.  CG can
behave badly if its step length ($\alpha$ in Alg.~\ref{alg:CG}) is
incorrect.  Figure \ref{F:strong_scaling:neg_pos:cg} shows the
results of solving the Poisson problem using \Code{FGmres->Cg->MueLu},
with both positive and negative SDC scaling factors.  We see that the
errors we deem ``bad'', i.e., permuting and scaling, are large.  For
GMRES, the figure is a symmetric version of
Fig.~\ref{F:strong_scaling:gmres:TpetraScalingProblem:32} (omitted due
to space).

\section{Weak Scaling Results} \label{S:weak_scaling}
We now take the Poisson problem and fix the work per processor to 100k unknowns.
We then weakly scale the problem to 24 process (one node) and 1032 processes (43 nodes) on the NERSC
Hopper cluster.  This yields a global matrix size of approximately 102.3
million and required approximately 1TB of memory.

\begin{figure}[htp]
  \centering
  \begin{subfigure}[t]{\columnwidth}
    \setlength{\abovecaptionskip}{\SingleColumnAboveCaption}
    \centering
    \includegraphics[width=\columnWidthFactor\columnwidth]{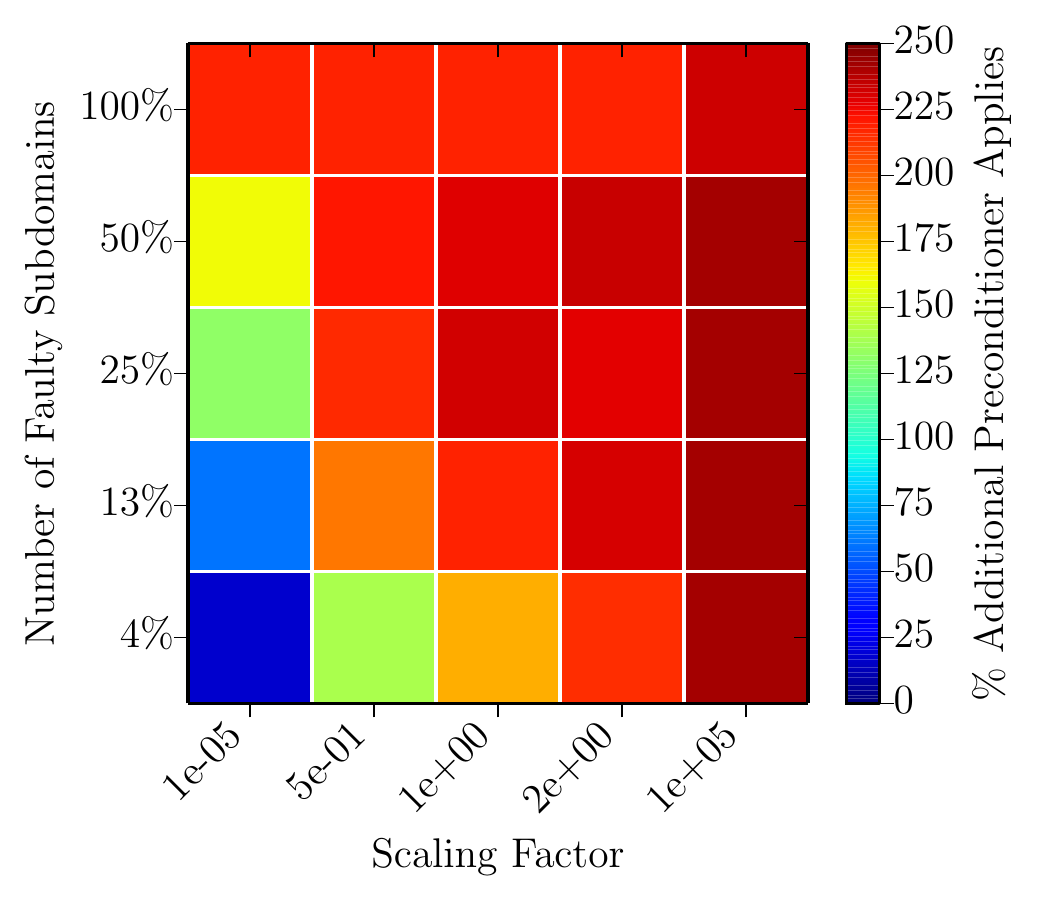}
    \caption{24 subdomains with 100k unknowns per
    process.}
    \label{F:weak_scaling:cg:24}
  \end{subfigure}%
  \\
  \begin{subfigure}[t]{\columnwidth}
    \setlength{\abovecaptionskip}{\SingleColumnAboveCaption}
    \centering
    \includegraphics[width=\columnWidthFactor\columnwidth]{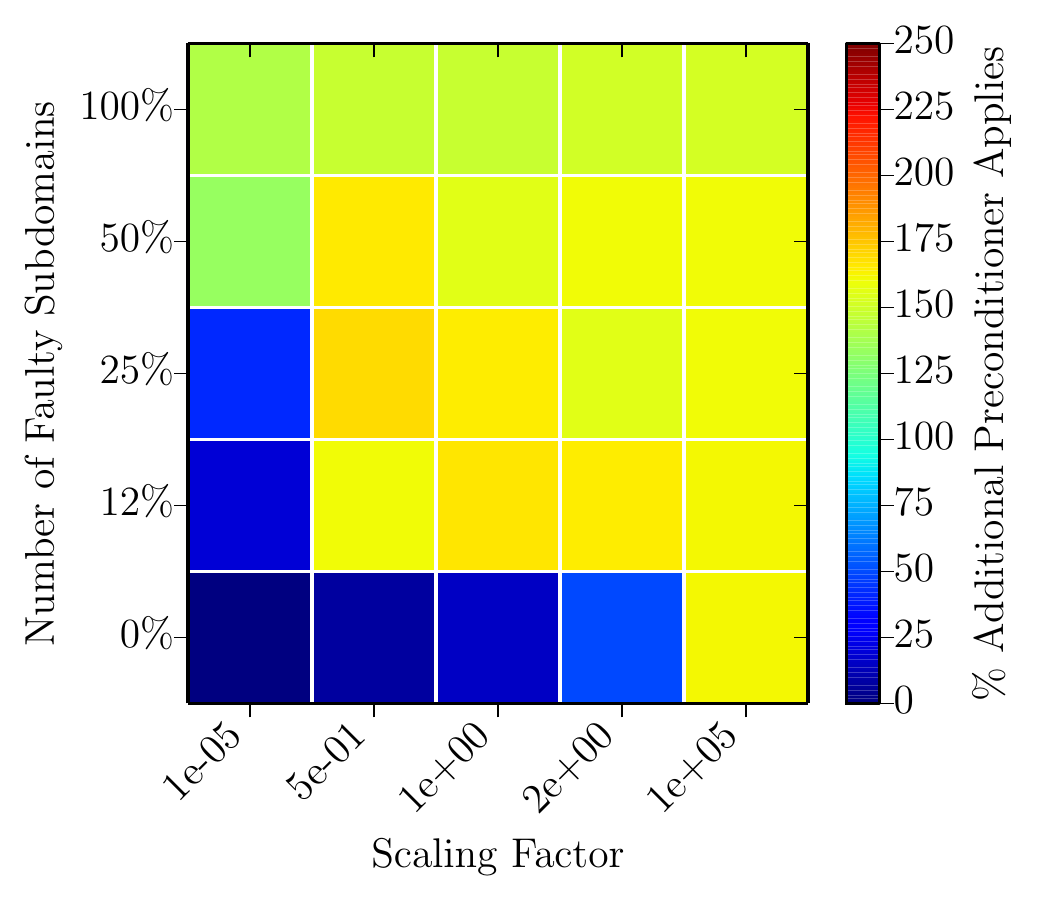}
    \caption{1032 subdomains  with 100k unknowns per
    process.}
    \label{F:weak_scaling:cg:1032}
  \end{subfigure}%
\setlength{\abovecaptionskip}{0.5\baselineskip}
\setlength{\belowcaptionskip}{-0.5\baselineskip}
  \caption{Weak Scaling (24 vs. 1032 subdomains): Average percent
  increase of preconditioner applies when solving an SPD problem using \Code{FGmres->Cg->MueLu}
  preconditioning weak scaled to 24 processors and 1032 processors}
  \label{F:weak_scaling:cg}
\end{figure}

Fig.~\ref{F:weak_scaling:cg} shows that the problem size has a clear impact on
fault tolerance. Should an error taint every subdomain, e.g., from
interpolations from a coarse grid, then we see nearly a 100\% different in
fault tolerance overhead between the worst SDC in the smaller problem
(Fig.~\ref{F:weak_scaling:cg:24}) and the larger problem
(Fig.~\ref{F:weak_scaling:cg:1032}).  We also see the repeated trend that faults
that maintain and do not drastically increase the 2-norm of the preconditioner
output typically incur lower fault tolerance overhead. Note: The bottom row of
squares (4\% and 0\% Faulty Subdomains) correspond to a single process creating
SDC.

\section{Conclusion} \label{S:conclusion}

This paper demonstrates that iterative linear solvers can get the
right answer despite incorrect arithmetic or storage in their
preconditioners.  They can do so without algorithmic or implementation
changes to preconditioners by combining selective reliability
and inner-outer iterations.  Fault detection in inner solves need not
catch all incorrect preconditioner results in order to reduce overhead
much below just running the solver twice.  This justifies even an
expensive implementation of reliability in outer solves, since most of
the work goes into inner solves with their more effective
preconditioner.  We have also shown that analytical approaches that
detect and filter out large errors scale well and significantly reduce
faults' overhead.  This is particularly true for effective
preconditioners like algebraic multigrid, that require only a few
solver iterations, but whose complexity and global communication
patterns may make them more vulnerable to SDC.  Combining the
projection norm bound from Elliott \cite{elliott14} with an occasional
check for monotonicity of the explicit residual norm in GMRES detected
more faults than either alone.

We have presented results based on a fault model that is possible
given future soft fault projections.  Given what little we know about
how faults will appear in future hardware, we have chosen to use
faults that represent an entire MPI process returning incorrect data.  Our
fault model does not aim to predict actual behavior of future SDC.
Rather, it shows a case sufficiently ``bad'' for us to assess how our
fault tolerance strategies behave when presented with very damaging
SDC.

Whether SDC turns out to be a real ``monster in the closet'' or not,
our findings are relevant for other fields of research.  We observe a
consistent trend in our data: Faults that increase the 2-norm are
worse than faults that maintain or decrease the 2-norm.  We also see
that when the number of faulty ranks is low, returning data that is
wrong but ``small'' (in the L-2 norm sense) is optimal.  This suggests
a strategy for dealing with failure of MPI processes or other data
loss, namely, replacing the missing data with ``small'' values and
continuing the solve.  Future work will pursue with our collaborators
this common strategy for recovery from both soft and hard faults.  We
also plan to compare the performance of this paper's approach with
that of software checksums and other resilience techniques.  Finally,
we will investigate the development and use of programming models that
provide selective reliability.

\section*{Acknowledgment}\label{S:ack}
This work was supported in part by grants from NSF (awards 1058779 and
0958311) and the U.S.\ Department
of Energy Office of Science, Advanced Scientific Computing Research,
under Program Manager Dr.\ Karen Pao.

Sandia National Laboratories is a multiprogram laboratory managed and
operated by Sandia Corporation, a wholly owned subsidiary of Lockheed
Martin Corporation, for the U.S.\ Department of Energy's National
Nuclear Security Administration under contract DE-AC04-94AL85000.

\bibliographystyle{IEEEtran}
\bibliography{paper}
\end{document}